\documentclass[aps,prl,preprint,onecolumn,superscriptaddress,dvipsnames]{revtex4-1}
\usepackage{amsmath,amssymb}
\usepackage{graphicx}
\usepackage{float}
\usepackage{natbib}
\usepackage{color}
\usepackage[colorlinks=true,bookmarks=false,citecolor=blue,urlcolor=blue]{hyperref}
\usepackage{lmodern}
\usepackage[most]{tcolorbox}
\usepackage{tikz}
\usepackage{varwidth}
\usepackage{wrapfig}
\usepackage{capt-of}
\usepackage{paracol}

\begin{document}

\title{Non-Planck thermal emission from two-level media}

\author{Igor A. Nechepurenko}
\affiliation{Moscow Institute of Physics and Technology, Dolgoprudny 141700, Russia}
\affiliation{Dukhov Research Institute of Automatics, Moscow 127055, Russia}

\author{Denis G. Baranov}
\email[]{denis.baranov@phystech.edu}
\affiliation{Center for Photonics and 2D Materials, Moscow Institute of Physics and Technology, Dolgoprudny 141700, Russia}

\begin{abstract}
Thermal emission is a universal phenomenon of stochastic electromagnetic emission from an object composed of arbitrary materials at elevated temperatures. A defining feature of this emission is the monotonic and rapid growth of its intensity with the object's temperature for most known materials. This growth originates from the Bose-Einstein statistics of the thermal photonic field. The fact that the material's ability to absorb and emit light may change with temperature, however, is often ignored.
Here, we carry out a theoretical study of thermal emission from structures incorporating ensembles of two-level media. We investigate this effect in a range of geometries including thin films and compact nanoparticles, and establish the general dependencies in the evolution of thermal emission from systems including two-level media. Thermal emission from such structures is essentially Non-Planckian and exhibits a universal asymptotic behavior in the limit of high temperatures.
These results might have important implications for the design of thermal energy harvesting and thermal vision systems.
\end{abstract}

\maketitle

\section{Introduction}

Any absorbing object radiates a broadband spectrum of electromagnetic radiation into the environment according to the principles of statistical mechanics.
The class of idealized black bodies -- objects that perfectly absorb any incident electromagnetic radiation -- emit broadband and incoherent thermal radiation, whose spectral radiance (spectral power density per unit area per unit solid angle radiated along the observation angle $(\theta,\varphi)$ is described by the Planck's law \cite{planck1901law}:
\begin{equation}
   L_{BB}=\frac{ \omega^2}{4\pi^3 c^2} \frac{\hbar \omega}{e^{\hbar \omega /k_B T}-1}.
   \nonumber
\end{equation}
Upon integration over the whole spectral and angular range it yields the Stefan-Boltzmann law describing the integral black-body radiation intensity \cite{stefan1879uber}:
\begin{equation}
   I_{BB}=\sigma T^4,
   \nonumber
\end{equation}
where $\sigma$ is the Stefan-Boltzmann constant.
As this law suggests, the integral intensity of emission from a black-body quickly grows with temperature. This rapid $T^4$ growth originates from the Bose-Einstein statistics of the population of the surrounding photonic field \cite{boyd1983radiometry}.

This dependence is modified for materials and nanostructures with selective absorption spectra, for which the spectral radiance $L$ is defined by the structure's absorptivity $\alpha(\omega)$. 
This correspondence between the emissivity and the absorptivity, known as the Kirchhoff's law \cite{Kirchhoff,greffet2018light}, has enabled a wide range of thermal emitters with narrowband \cite{Celanovic2005, Liu2011, NodaReview}, directional \cite{Greffet2002, Laroche2005, Focused}, and tunable \cite{Inoue2014, Du2017,Vanadium} emission spectra that have found numerous practical applications for thermal energy management \cite{Byrnes2014,davids2020electrical} and radiative cooling \cite{gentle2010radiative, Raman2014, mandal2018hierarchically, Zhai2017}, (see \cite{baranov2019nanophotonic} for a review).
Although the Kirchoff's law enables a slower growth of thermally emitted intensity with temperature from narrowband emitters, a more dramatic evolution of emitted intensity with temperature is needed in certain situations.
Finding a material or a structure exhibiting strongly non-Planckian thermal emission is of great importance for many practical applications dealing with thermal light, such as energy recycling \cite{Byrnes2014} and infrared imaging \cite{qu2018thermal}.
To achieve this, the emissivity of the material itself must vary with temperature.

Media and nanostructures whose ability to emit light (i.e., their emissivity) varies with temperature, form the set of \textit{thermochromic} materials.
There has been a considerable progress in studies on non-Planckian thermal emission from nanostructures made of thermochromic crystals such as GST \cite{Du2017,qu2017dynamic}, VO$_2$ \cite{Vanadium}, or SmNiO$_3$ \cite{shahsafi2019temperature}.
An interesting case of thermochromics is presented by so called negative differential emission, wherein the total thermal emission intensity does not grow, but in contrast, reduces upon heating in a certain interval of temperatures \cite{Vanadium}. An extreme regime is the zero-differential, or invariant, thermal emission - the regime wherein the total intensity of emission from an object remains (nearly) constant in a range of temperatures \cite{shahsafi2019temperature}. When viewed through a thermal camera, such an object would appear to maintain its temperature even when heated or cooled within that range of temperatures.



In this paper, we theoretically investigate thermal emission from structures incorporating ensembles of two-level emitters.
Due to the statistical behavior of the two-level medium population, its emissivity quickly falls with temperature, thus rendering the resulting thermal emission non-Planckian.
We investigate thermal emission in a range of common geometries including thin films and compact nanoparticles, and find the general dependencies in the evolution of thermal emission from systems incorporating two-level media.  Furthermore, we demonstrate that within the framework of a two-level medium, the intensity of emitted thermal radiation universally approaches a constant value in the limit of high temperatures for any finite structure.
We expect these results to be valuable for the development of energy recycling systems and thermal camouflage.

\begin{figure}[t!]
\centering\includegraphics[width=0.8\columnwidth]{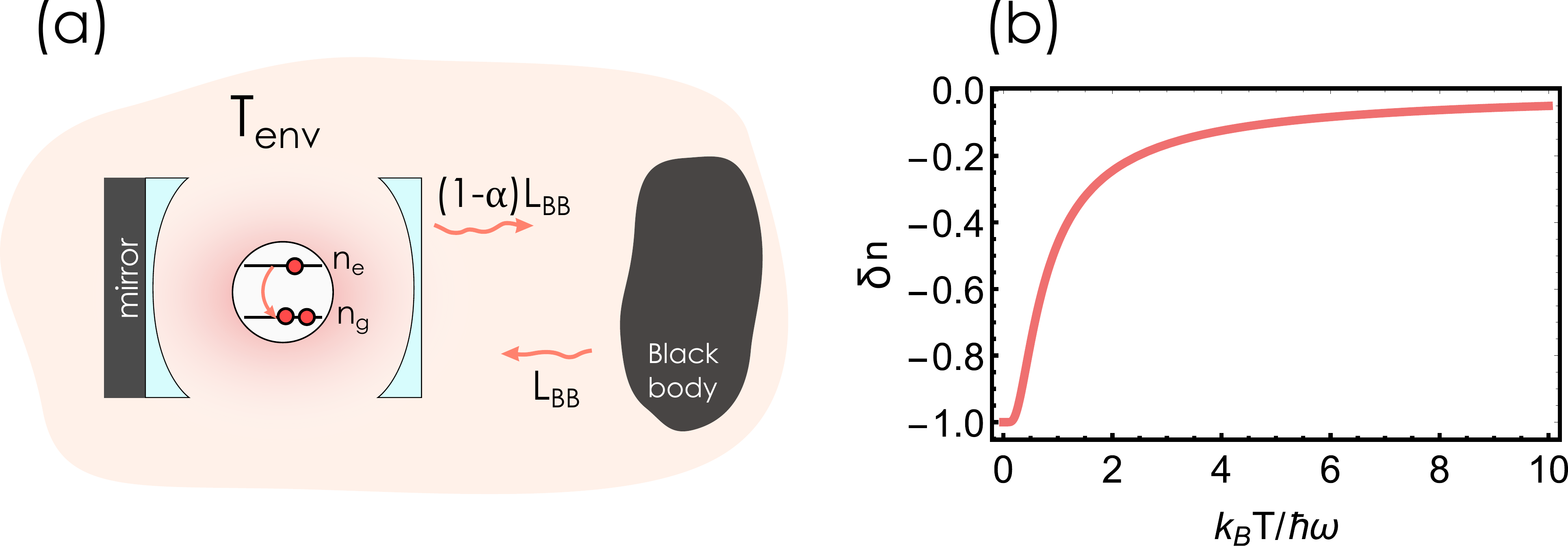}
\caption{
(a) Sketch of a thermal emitter incorporating a medium formed by two-level emitters. The system exchanges electromagnetic energy with the environment at temperature $T$. (b) Temperature-dependent equilibrium population inversion $\delta n$ of a two-level medium.}
\label{concept}
\end{figure}

\section{Results}

Ensemble of two-level emitters forming a continuous medium is a commonly used generic system in studies of thermal emission.
Fig. 1(a) illustrates the process of thermal emission from such a system: a generic cavity enclosing a two-level medium exchanges energy with the surrounding photonic field at temperature $T$. As a result of this exchange, the structure emits a continuous spectrum of thermal radiation.
The integrated thermal emission intensity in the far field from a structure characterized by an absorptivity $\alpha$ has the form:
\begin{equation}
    I=\int_0^\infty {\alpha(\omega) L_{BB} d\omega},
\end{equation}
where $L_{BB}$ is the black-body radiance and $\alpha$ is the structure absorptivity spectrum. 

Dielectric properties of a two-level medium can be described by the Lorentzian permittivity:
\begin{equation}
    \varepsilon(\omega)=\varepsilon_{\infty}- f \delta n \frac{\omega_0^2}{\omega_0^2 - \omega^2 - i \gamma \omega},
\end{equation}
where $\varepsilon_{\infty}$ is the non-resonant permittivity of the medium, $\omega_0$ and $\gamma$ are the resonance frequency and linewidth, $f$ is the dimensionless bulk oscillator strength, and $\delta n = n_e - n_g$ is the ensemble average population inversion of the TLS. One often overlooked detail is that the population inversion of the two-level medium is a function of temperature itself. At a zero temperature all emitters are in their ground state and  $\delta n = -1$. Applying the Gibbs distribution to an ensemble of two-level systems at temperature $T$, we find the thermal population:
\begin{equation}
\delta n = \frac{e^{-\hbar \omega_0/k_B T}-1}{1+e^{-\hbar \omega_0/k_B T}}.
\end{equation}
The population inversion becomes less negative with temperature, Fig. 1(b), thus effectively making the material less absorbing, hence suppressing its thermal emittance. This dependence readily allows us to calculate the absoprtivity and integrated thermal emission intensity of from an arbitrary two-level medium at a temperature $T$.


To begin our analysis of thermal emission from structures incorporating two-level media, we consider thermal emission from a homogeneous film described by permittivity of the form Eq. (2) with $\varepsilon_{\infty}=6$, $f=0.4$, $\gamma / \omega_0 = 0.05$  in air. Fig. 2(a) shows the resulting emissivity $\epsilon(\omega)$ of the film for a series of temperatures $T/\omega_0$ (we use dimensionless temperature $k_BT/\hbar$ in the plots). These spectra clearly indicate that the emissivity of the film made of a two-level medium quickly drops with temperature  as a result populating the excited level of the emitters, Fig. 1(b).

\begin{figure}[hbt!]
\includegraphics[width=\columnwidth]{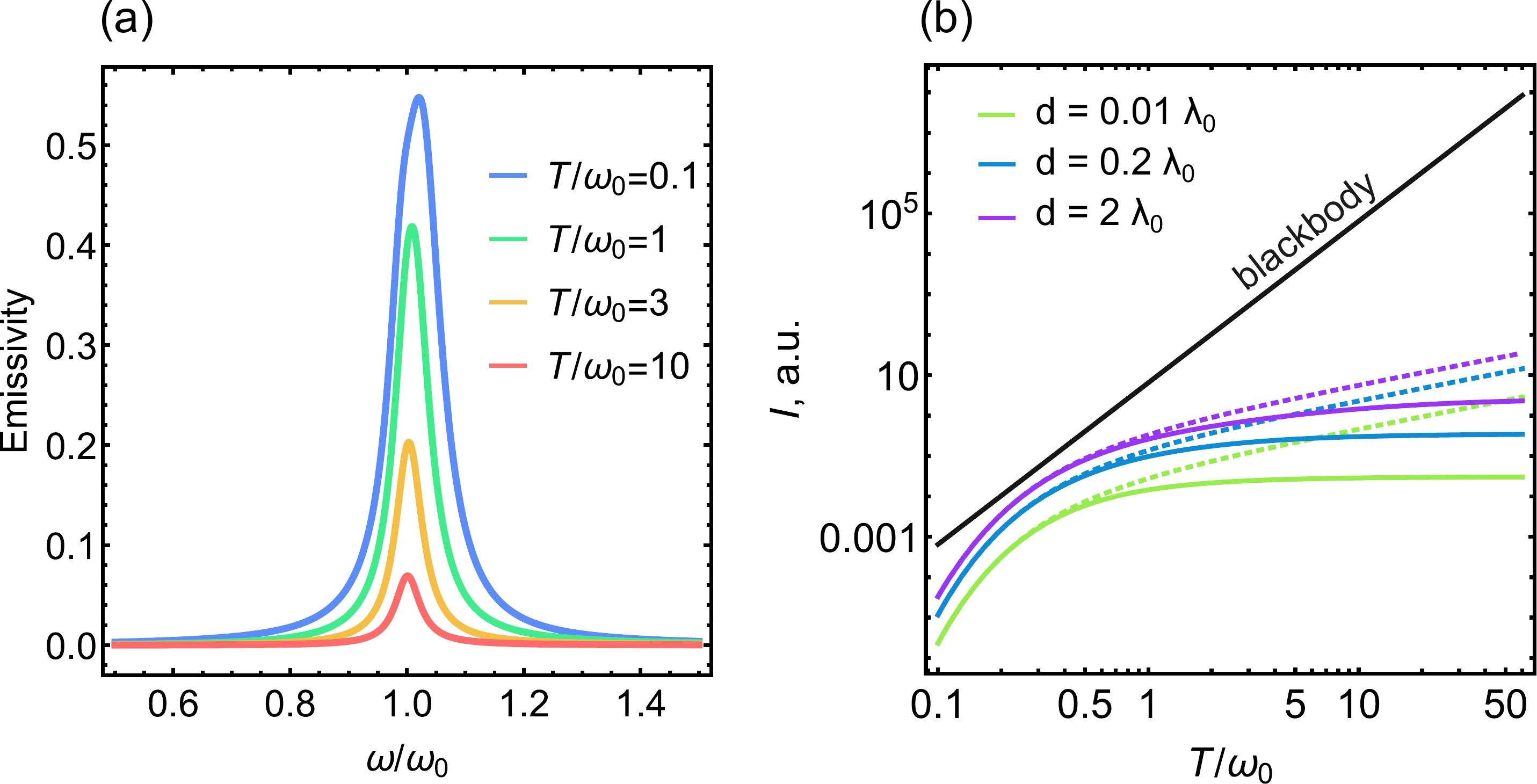}
\caption{(a) Temperature-dependent emissivity of a thin film of thickness $d=\lambda_0/10$ made of a two-level medium described by the Lorentz permittivity $\varepsilon(\omega,T)$, Eq. (2), in vacuum. (b) Temperature-dependent integral intensity of thermal radiation $I$ emitted in the normal direction from a thin film made of a two-level medium for a series of thickness values $d$. The black line is the blackbody emission intensity; dashed lines show the corresponding emission intensity from a harmonic Lorentz film of the same thickness.}
\label{concept}
\end{figure}

Fig. 2(b) presents integral total emission intensity $I$ (radiated normally to the interface) as a function of temperature for a film of thickness $d$ in free space in comparison to the intensity emitted by a black-body $I_{BB}$. It readily shows that the emission intensity radiated by a film of two-level medium saturates at temperatures $k_B T \approx \hbar \omega_0$ irrespective of the film thickness and approaches a constant in the limit of high temperatures. Correspondingly, due to this saturation it starts to lag behind the black-body emission $I_{BB}$ at temperatures above the characteristic temperature of the two-level medium.

Interestingly, the intensity of thermal emission from two-level films features a super-Planckian growth at low temperatures, $k_B T < 0.3 \hbar\omega_0$, Fig. 2(b). This is the result of the blackdoby emission peak approaching the resonant frequency of the two-level medium: the population inversion is nearly stationary at low temperatures, but the magnitude of the Planck's blackbody factor $L_{BB}$ grows rapidly in the vicinity of $\omega_0$ (with the blackdoby emission peak approaching the resonance from the low-energy side). Since most of the emitted intensity originates from energies close to $\omega_0$, this leads to super-Planckian growth of the integral intensity.
This observation might be interesting in the context of super-resolution based on the super-linearity of thermal radiation \cite{graciani2019super}.

It is instructive to compare the resulting thermal emission from a two-level medium film with the case of a harmonic Lorentz medium, i.e., a medium formed by an ensemble of resonant harmonic oscillators, rather than two-level transitions. Because of a different statistical behavior of the harmonic oscillator, its absorptivity (and emissivity, respectively) does not get saturated at elevated temperature and remains constant: $\delta n = -1$. The resulting integral emission intensity of harmonic Lorentz films grows slower than the blackbody intensity, but does not remain constant at high temperatures (dashed curves in Fig. 2(b)). Fitting the resulting intensity in the double logarithmic scale yields linear dependence at high temperatures, $I \propto T$. This scaling can be understood by noting that the emissivity of a harmonic film $\epsilon(\omega)$ does not vary with temperature; at the same time, at high temperatures $L_{BB} \propto T$ at a fixed frequency, resulting in the linear growth of the integral emission intensity.

As another representative example of emitting structure, we consider a thin film of a two-level medium of thickness $L$ placed a distance $h$ from a lossless metallic mirror.
The resulting dependence of the integral thermal emission radiated at a normal from the two-level medium film exhibits a behavior very similar to that observed with a thin film in free space, Fig. 3(a).

\begin{figure}[hbt!]
\includegraphics[width=\columnwidth]{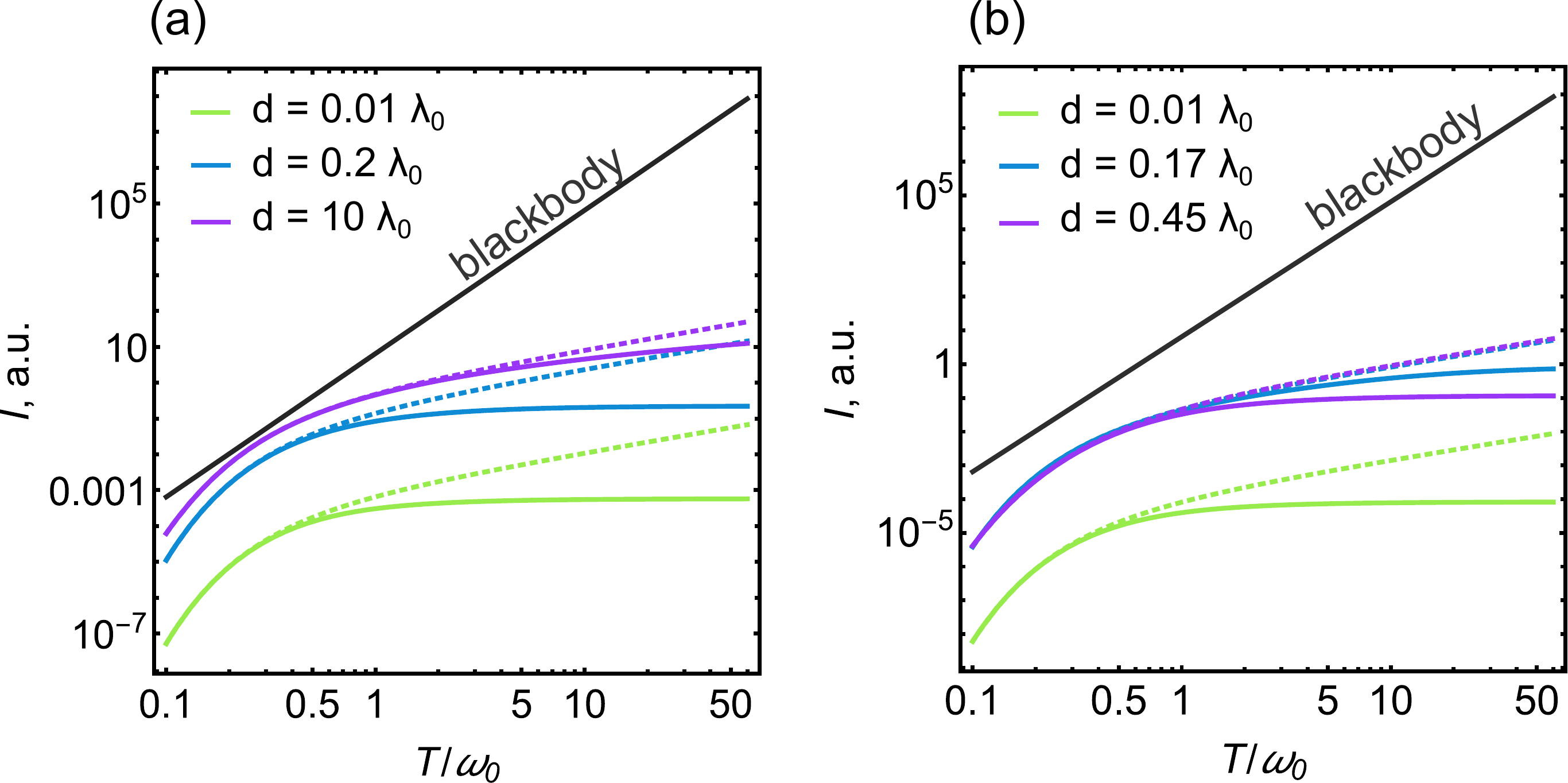}
\caption{(a) Temperature-dependent integral intensity of thermal radiation $I$ emitted in the normal direction from a two-level medium film of thickness $d$ placed on top of a perfect electric conductor. The dashed line is the corresponding blackbody emission intensity;  (b) Temperature-dependent intensity of thermal emission from a two-level medium film of thickness $d$ placed between a mirror with 90\% reflectance and a perfect electric conductor. The black line is the blackbody emission intensity; dashed lines show the corresponding emission intensity from a harmonic Lorentz film of the same thickness in the same geometry.}
\label{fgi3}
\end{figure}

Next, we consider a similar two-level medium film, but enclosed in a Fabry-Perot cavity formed by another semi-transparent mirror on top (described by a lossless model $\varepsilon_D = 1-\omega_p^2/\omega^2$ with $\omega_p=12\omega_0$ 
chosen for the demonstration). The cavity can selectively enhance thermal emission at its resonant wavelengths, thus decreasing the bandwidth of thermal emission from a resonant medium even further \cite{Celanovic2005}. The resulting dependencies of integral emission intensity shown in Fig. 3(b) demonstrate a behavior qualitatively similar to other thin film geometries examined above - the integral intensity approaches a constant at high temperatures, whereas emission intensity from a harmonic Lorentz film grows as $I \propto T$. Notably, thicker cavities filled with the two-level medium can exhibit lower emission intensities, as we can infer from comparing the curves corresponding to $d=0.17 \lambda_0$ and $d=0.45 \lambda_0$. This behavior originates from the critical coupling of cavities of certain thickness, resulting in maximized absorption and emissivity.



Finally, we examine non-Planck thermal emission from compact objects incorporating two-level absorbing media. As a tutorial example we consider a solid sphere of radius $r$ made of a two-level material described by the permittivity Eq. (2). For the case of a compact object, the integral power of thermal radiation emitted in the observation direction $(\theta ,\varphi)$ per unit area of the object is determined by the corresponding absorption cross-section $\sigma_{(\theta,\varphi)}$ \cite{BiehsPRB,sakat2018enhancing}:
\begin{equation}
    I = \frac{1}{S} \int_0^\infty { \sigma_{(\theta,\varphi)}(\omega) L_{BB} d\omega},
\end{equation}
where $S$ is the geometrical cross-section.
Fig. 4 presents the resulting thermal power per unit area emitted by a sphere for two illustrative cases: a subwavelength sphere with $\varepsilon_{\infty} = 1$, whose response is dominated by the Frolich (plasmon-like) resonances occurring at negative permittivities, and a wavelength-scale sphere with $\varepsilon_{\infty} = 6$, whose response is dominated by dielectric Mie resonances coupled with the Lorentz resonance of the two-level medium. Despite the difference, both situations feature qualitatively similar dependencies of thermal emission intensity, saturating near $k_B T \approx \hbar \omega_0$.

\begin{figure} 
\includegraphics[width=\columnwidth]{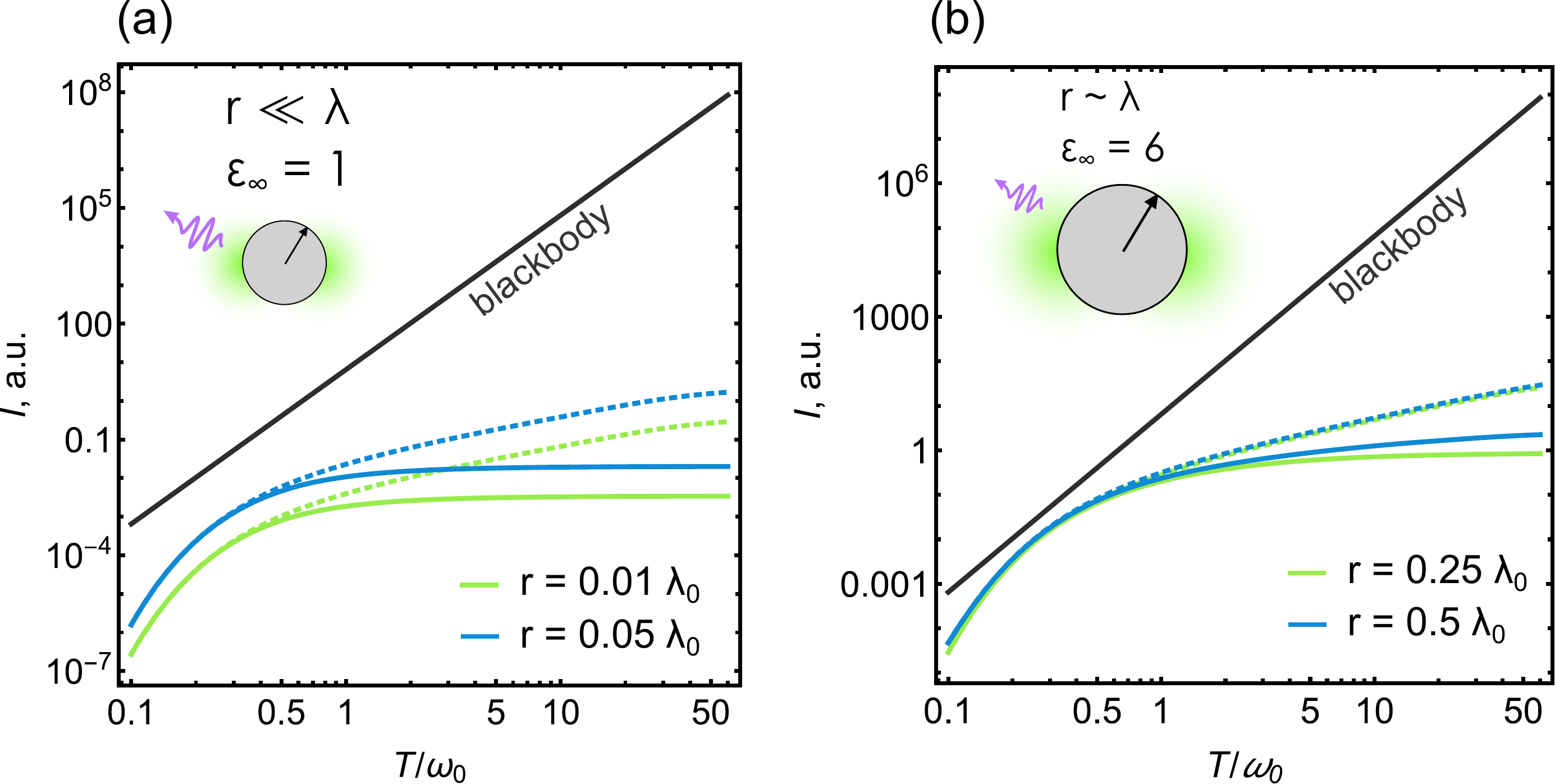}
\caption{(a) Temperature-dependent integral intensity of thermal radiation $I$ emitted from a subwavelength sphere of radius $r$ in air made of two-level medium. The dashed lines show the corresponding emission intensity from a sphere made of the harmonic Lorentz material;  (b) The same as (a) for a wavelength-scale sphere described by the same permittivity with $\varepsilon = 6$.}
\label{fgi3}
\end{figure}

It is possible to build an analytical argument showing that in the limit of high temperatures the integral emission intensity from an arbitrary finite structure incorporating only a perfect two-level medium as a source of thermal fluctuations approaches an asymptote.
Absorptivity $\alpha$ (or alternatively scattering cross-section $\sigma$ for a compact object) of the system under excitation with a plane wave $\mathbf{E}_{inc} = \mathbf{E}_{0}e^{i \mathbf{k}_{inc} \mathbf{r}}$ from a certain direction $(\theta,\varphi)$ can be found as work performed by the total field $\mathbf{E}_{tot}$ on the induced electric current density $\mathbf{j}(\mathbf{r})$:
\begin{multline}
    \alpha (\omega,T) \propto \frac{1}{I_{inc}}
    \int _V \frac{1}{2} \operatorname{Re} ( \mathbf{E}_{tot}(\mathbf{r}) \cdot \mathbf{j}^*(\mathbf{r}) ) d^3 \mathbf{r} = \\
  \frac{\omega }{2 I_{inc}} \int _V \varepsilon_0\ \operatorname{Im} \varepsilon(\mathbf{r},\omega,T) |\mathbf{E}_{tot} (\mathbf{r},\omega,T) |^2 d^3 \mathbf{r}.
\end{multline}
where $I_{inc}$ in the incident energy flux. Expanding the imaginary part of the permittivity in the series over $1/T$, we obtain
\begin{equation}
    \operatorname{Im} \varepsilon(\omega,T) = A\frac{1}{T} + o(1/T),
\end{equation}
where $A$ is a constant. 
The total field can be expanded in a similar way. In the limit of high temperatures the population inversion $\delta n$ approaches zero, and the material becomes non-resonant.
Assuming the limiting structure ($\delta n = 0$) does not supports bound states in the continuum, the induced electric field across the structure $\mathbf{E}_{tot} (\mathbf{r},\omega)$ can be expanded as
\begin{equation}
    \mathbf{E}_{tot}(\mathbf{r},\omega,T) =  \mathbf{E}_{\infty}(\mathbf{r},\omega) + \delta \mathbf{E}(\mathbf{r},\omega,T) O(1/T),
\end{equation}
where $\mathbf{E}_{\infty}(\mathbf{r},\omega)$ is the limiting field distribution determined at every wavelength and illumination direction $(\theta,\varphi)$ solely by the geometry and $\varepsilon_\infty$.
Substituting this into Eq. (5) we obtain $\alpha (\omega,T) \propto 1/T$ in the limit of high temperatures.
Finally, combining this absorptivity with the black-body radiance $L_{BB}$, integrating within the linewidth of the emitter $\gamma$ and keeping only the leading term in the expansion, we find the asymptotic behavior of the total thermal emission:
\begin{equation}
    I_{TLS} = \int_0^\infty {\alpha(\omega,T) L_{BB}(\omega,T) d\omega} = 
    C + O(1/T),
\end{equation}
where $C$ is a constant.

This is a universal result that holds for any ensemble of truly two-level systems. Unfortunately, this saturation behavior is quite slow and becomes pronounced only at temperatures above the transition frequency of the resonant medium, $k_B T>\hbar \omega_0$, which makes this effect not suitable for realization of temperature-independent thermal emission \cite{shahsafi2019temperature}. However, we expect that combining this behavior with other thermochormic materials, for example, with multi-layered structures incorporating dielectric layers with temperature-dependent refractive indices, could shift the temperature-independent emission regime to lower temperatures.

We realize that a two-level medium is an idealization, and any real material inevitably will contain higher-energy vibrational or electronic transitions associated to respective resonances of the material's permittivity. However, even if one accounts for these transitions, they will contribute to spectral radiance only in the frequency range near the respective resonance. Therefore, if the quantity of interest is the emission intensity integrated within a certain wavelength range (which is the case, for example, for most of thermal imaging devices), the resulting behavior of this quantity will exhibit a saturation similar to the ones demonstrated above. 

\section{Conclusion}
To conclude, we have studied temperature dependencies of thermal emission from structures incorporating resonant two-level media.
We have examined a range of practically common geometries such as thin films, cavities, and spherical nanoparticles, 
and found that thermal emission from all these systems shares one common feature: its intensity is non-Planckian, and approaches a constant value (dependent on the structure) in the limit of high temperature. This asymptotic behavior stems from the trade-off between the diminishing emissivity of the two-level medium, and the growing population of the photonic field.
These results might have far reaching implications for the design of thermal energy harvesting and thermal vision systems.

\section{Acknowledgments}
The work was supported by the Russian Science Foundation (21-12-00316).
D.G.B. acknowledges support from the Russian Federation President Grant (MK-1211.2021.1.2).

\bibliography{thermal}

\end{document}